# Universally enhanced superconductivity and coexisting ferroelectricity at oxide interfaces

Meng Zhang[1,5,*], Ming Qin[1,5], Yanqiu Sun[1,5], Siyuan Hong[1] and Yanwu Xie[1,2,3,4,*]

[1]School of Physics, and State Key Laboratory for Extreme Photonics and Instrumentation, Zhejiang University, Hangzhou 310027, China
[2]College of Optical Science and Engineering, Zhejiang University, Hangzhou 310027, China
[3]Collaborative Innovation Center of Advanced Microstructures, Nanjing University, Nanjing 210093, China
[4]Hefei National Laboratory, Hefei 230088, China.
[5]These authors contributed equally: Meng Zhang, Ming Qin, Yanqiu Sun.

*e-mail: physmzhang@zju.edu.cn; ywxie@zju.edu.cn

**Abstract**

The coexistence of superconductivity and ferroelectricity is rare due to their conflicting requirements: superconductivity relies on free charge carriers, whereas ferroelectricity typically occurs in insulating systems. At LaAlO$_3$/KTaO$_3$ interfaces, we demonstrate the coexistence of two-dimensional superconductivity and ferroelectricity, enabled by the unique properties of KTaO$_3$ as a quantum paraelectric. Systematic gating and poling experiments reveal a universal enhancement of the superconducting transition temperature ($T_c$) by ~0.2–0.6 K and bistable transport properties, including hysteresis, strongly suggesting the existence of switchable ferroelectric polarization in the interfacial conducting layer. Hysteresis loops indicate robust ferroelectricity below 50 K. The $T_c$ enhancement is attributed to ferroelectric polarization-induced reduction in dielectric constant, which narrows the interfacial potential well, confining carriers closer to the interface. The bistability arises from switchable ferroelectric polarization, which modulates the potential well depending on polarization direction. These findings establish a straightforward mechanism coupling ferroelectricity and superconductivity, providing a promising platform for exploring their interplay.

# Main

The coexistence of superconductivity and ferroelectricity represents a long-standing challenge due to their fundamentally conflicting requirements. Superconductivity demands on a high density of free charge carriers, while ferroelectricity is generally found in insulating materials. The introduction of free charges into such materials screens long-range Coulomb interactions, suppressing ferroelectricity. Although the coexistence of free carriers and polarization has been proposed[1] and observed[2,3] in polar metals, including polar superconductors[4,5], the polarization in these systems is non-switchable. Recently, however, superconductivity and ferroelectricity were observed to coexist in two-dimensional (2D) van der Waals heterostructures[6], where ferroelectricity arises from mechanisms distinct from conventional Coulomb interactions.

Oxide interfaces, particularly those based on SrTiO$_3$ (STO)[7–9] and KTaO$_3$ (KTO)[10–15], offer a compelling alternative for exploring the interplay between superconductivity and polarization. Both STO and KTO are wide-gap semiconductors (3.2 eV for STO and 3.6 eV for KTO) and quantum paraelectric[16,17]. Electron-doped STO was the first oxide superconductor discovered[18], and its proximity to ferroelectricity inspired the discovery of high-temperature cuprate superconductors[19]. Quantum ferroelectric fluctuations have been proposed as a potential pairing mechanism in these systems[20–27], and a ferroelectric quantum phase transition was observed inside the superconducting dome of STO[28]. Additionally, when STO is combined with insulating oxides such as LaAlO$_3$ (LAO), 2D superconductivity emerges at the interface[8]. Recent studies have shown that switchable 2D electron gas[29] and 2D superconductivity in ferroelectric-like environment[19] can be achieved in STO-based interfaces via methods such as Ca alloying or $^{18}$O substitution. However, the coexistence of superconductivity and switchable ferroelectric polarization remains elusive in STO-based systems.

In contrast, while superconductivity has not been observed in electron-doped bulk KTO[30], recent studies reveal that 2D superconductivity can emerge at KTO surfaces[31,32] and interfaces[11–13]. Notably, KTO-based systems exhibit significantly higher superconducting transition temperature ($T_c$) than STO-based interfaces, making them more promising for exploring the interplay between ferroelectric polarization and superconductivity. Here, we demonstrate the coexistence of superconductivity and ferroelectricity at LAO/KTO interfaces, providing a novel platform for studying this unconventional phenomenon.

**Universally enhanced superconductivity and bistability**

A noteworthy feature of KTO interface superconductivity is its tunability via gate bias ($V_G$) across KTO[13,33–35], as illustrated in Fig. 1a. Over the past several years, we have gated more than 100 LAO/KTO samples and identified a universal phenomenon: after gating experiments (where a $V_G$ of up to ±200 V was applied and subsequently removed, referred to as "poling"), the $T_c$ of LAO/KTO exhibited a significant enhancement of ~0.2-0.6 K (see Fig. 1b for selected results). Additionally, the normal-state sheet resistance ($R_s$) exhibits a bistable characteristic after poling with different bias polarities (Fig. 1c), indicating the presence of ferroelectricity.

**Ferroelectric hysteresis under gating cycles**

To further investigate these phenomena, we examined the transport behaviors of a typical LAO/KTO(111) Hall bar device under continuous gating cycles. For each $V_G$ value, both the temperature dependence of $R_s(T)$ and the Hall effect (measured at $T = 4.5$ K) were recorded. Figure 2a shows the $R_s(T)$ curve before any $V_G$ was applied (denoted as "origin"). A metallic behavior, followed by a superconducting transition at $T_c = 1.89$ K (defined as the temperature where $R_s$ drops to 50% of its normal-state value at 4.5 K), was observed. Figures 2b and 2c display the $R_s(T)$ curves during a cycle where $V_G$ was swept from +180 V to -180V and then back. In both sweeping directions, a clear overall tuning effect consistent with previous studies[13,33] was observed: a positive (negative) $V_G$ decreases (increases) the normal-state $R_s$ and lowers (raises) $T_c$. However, in addition to this overall tuning effect, the $R_s(T)$ curves exhibit a strong dependence on the gating history. For example, the three different $V_G = 0$ states ("origin", "$0^+$", and "$0^-$", where "$0^+$" and "$0^-$" represent the $V_G = 0$ state after removing positive or negative $V_G$, respectively) yielded significantly different $R_s(T)$ curves.

As summarized in Figs. 2d-f, two notable and correlated features emerge beyond the overall tuning effect. First, during the initial run starting from the "origin" state (indicated by the blue lines), the device undergoes an irreversible change likely associated with polarization formation, corresponding to the universal $T_c$ enhancement. Second, after the initial run, the device exhibits repeatable and pronounced hysteresis in both $V_G$-$R_s$ (Fig. 2d) and $V_G$-$T_c$ (Fig. 2e) loops. Analysis of the $V_G$-$R_s$ loops at different temperatures, measured during both decreasing (Fig. 2g) and increasing (Fig. 2h) temperature orders, reveals that hysteresis begins above 50 K and becomes pronounced below 30 K, consistent with the reported temperature at which dipole moments form in KTO[36–38].

**Effects of poling magnitude, temperature, and time**

Further investigations of the $0^\pm$ states after poling with $V_G$ (denoted as $V_G^{0\pm}$) were conducted to gain deeper insights into the observed ferroelectric behaviors. Since LAO/KTO is sensitive to gating history, we ensured an "origin" state in different experiments by using fresh samples when necessary. Figures 3a-d show the effects of poling $V_G$ polarity and magnitude. Two identical Hall bar devices (S2_A# and S2_B#) cut from the same LAO/KTO sample were used. Similar experiments were also conducted on a single device (S7#, Extended Data Fig. 1), whose "origin" state was regenerated through a refreshing process (described below).

For each polarity, the $V_G$ was gradually increased from 0 to ±180 V. At each $V_G$, we applied the poling $V_G$ at 4.5 K for 3 minutes, followed by $R_s(T)$ and Hall effect measurements after removing $V_G$ to 0. As shown in Figs. 3a-b, and Extended Data Fig. 1a, the $T_c$ increases with the magnitude of poling $|V_G|$ for both polarities, and $T_c$ for "$0^-$" is consistently higher than for "$0^+$". Poling up to ±180 V, corresponding to ±3.6 kV/cm, did not saturate $T_c$ enhancement, suggesting that the ferroelectric polarization remains unsaturated. As shown in Figs. 3c-d, and Extended Data Fig. 1b, the increase in $T_c$ was accompanied by a clear decrease in $\mu$ (with a slight increase in $n_s$; however, the change in $n_s$ was much smaller than that of $\mu$). This indicates that, as discussed below, the ferroelectric polarization primarily modulates the interfacial potential well rather than directly altering $n_s$.

Poling temperature also plays a critical role. As shown in Fig. 3e, $T_c$ begins to increase sharply when poling temperatures dropped below 30 K, matching the temperature at which $V_G$-$R_s$ hysteresis becomes pronounced (Figs. 2g-h). Notably, significant $T_c$ enhancement was observed after just one second of poling, with saturation occurring after 10-15 seconds of cumulative poling time (Fig. 3f and Extended Data Fig. 2). The tuned states remain nonvolatile at low temperatures (Fig. 3g). Full recovery to the "origin" state can be achieved by leaving the samples at ambient conditions for several days (Fig. 3h and Extended Data Fig. 3).

**Induced ferroelectricity in KTO**

All these experimental observations support that ferroelectricity emerges in the KTO side of LAO/KTO interfaces. This is unsurprising, as ferroelectricity has previously been suggested in KTO through minimal chemical doping (*e.g.*, Nb, Li)[39], oxygen deficiency[40], strain[41], or randomly distributed defects[37,42,43]. Given the relatively low ferroelectric onset temperature observed in LAO/KTO, defect-induced polar clusters likely play a critical role[37,43]. Although ionic migration could theoretically result in switchable polarization, such migration is typically slow and occurs predominantly at high temperatures[44–46], which is inconsistent with our experimental findings.

**Coexistence of ferroelectricity and superconductivity in the conducting layer**

As illustrated in Fig. 1a, superconductivity at KTO interfaces is confined within a thin KTO layer (~5-10 nm thick[11–13,33]), corresponding to the width of the interfacial potential well. The remaining KTO bulk serves as a thick insulator across which $V_G$ is applied. While ferroelectric polarization is expected in the insulating bulk, we propose that it also occurs in the conducting layer. If polarization were confined solely to the insulating bulk, as in conventional ferroelectric transistors, its effects would primarily manifest in $n_s$ rather than $\mu$, inconsistent with our observations. Moreover, the bistability in transport properties suggests a switchable polarization within the conducting layer, whose built-in field modulates the interfacial potential well. These results lead us to conclude that ferroelectric polarization coexists with 2D superconductivity in the interfacial conducting layer. Although ferroelectricity and conductivity are traditionally considered mutually exclusive, the polarization in the conducting KTO layer may arise from coupling with ferroelectricity in the insulating KTO bulk due to their inherent lattice bonding effects.

**Ferroelectric polarization modulating the interfacial potential well**

We now address how the presence of ferroelectricity in the conducting KTO layer can explain our experimental observations. Before delving into the specific phenomena associated with ferroelectricity, we reinforce that the transport properties of LAO/KTO are largely governed by the interfacial potential well profile, with $\mu$ serving as a key indicator. Previous studies[13,33,34] have shown that, particularly when $n_s$ is relatively large, gating at KTO interfaces primarily modulates $\mu$ rather than $n_s$. As shown in Fig. 2f and Extended Data Fig. 4a, sweeping $V_G$ from -180 V to 180 V caused $\mu$ to vary from ~20 cm² V⁻¹ s⁻¹ to ~90 cm² V⁻¹ s⁻¹, while $n_s$ changed only slightly (and in this device, in a manner opposite to that expected from a simple capacitance effect. See Extended Data Figs. 4b-d for more information). This behavior can be attributed to $V_G$-induced modulation of the interfacial potential well, which alters the spatial distribution of carriers[13,47,48]. Such modulation influences the "effective disorder", thereby affecting $\mu$ (a narrower potential well reduces $\mu$)[13,47,48].

Notably, the hysteresis observed in $V_G$-$R_s$ (Fig. 2d) and $V_G$-$T_c$ (Fig. 2e) loops is mirrored in $V_G$-$\mu$ loops (Fig. 2f) but not in $V_G$-$n_s$ loops (Extended Data Fig. 4a). Furthermore, the universal $T_c$ enhancement is accompanied by a significant decrease in $\mu$ (Figs. 2f, 3c-d, and Extended Data Fig. 1b), which can be attributed to narrowing of the interfacial potential well[13,33,34]. Therefore, in LAO/KTO interfaces, unlike typical ferroelectric effects, ferroelectricity primarily affects transport by modulating the interfacial potential well profile rather than directly altering $n_s$.

Ferroelectric polarization in the interfacial KTO layer has two key effects: (1) it reduces the dielectric constant (ε) (Fig. 4a) and (2) introduces a built-in electric field due to polarization and the associated screening charges (insets of Fig. 4b). KTO, as an quantum paraelectric, exhibits a large ε of up to 4500 at low temperatures[17]. However, ε decreases dramatically under applied fields[49–51] and induced polarization[36,38]. As illustrated in Fig. 4a, after poling, ferroelectric polarization, regardless of its direction, lowers ε[36,38], thereby narrowing potential well (solid blue and red lines vs. dashed line, Fig. 4b), which explains the universal $T_c$ enhancement. Bistability naturally arises from switchable polarization, which can either narrow or widen the potential well depending on its direction (insets of Fig. 4b).

As shown in the $V_G = 0$ region of Fig. 2e, the $T_c$ enhancement after initial run is over 3 times larger than the $T_c$ difference between the "$0^+$" and "$0^-$" states in subsequent runs. This indicates that the reduction in ε due to polarization has a stronger influence on the potential well profile than built-in field of the polarization alone. The higher $T_c$ in the "$0^-$" state compared to the "$0^+$" state can be attributed to polarization narrowing the potential well in the former and widening it in the latter. The inability to fully recover the origin state by warming the samples above the ferroelectric onset temperature (even to room temperature; see Extended Data Fig. 3a) is likely due to unbalanced screening charges[52,53] at the insulating LAO surface persisting in the cryostat environment. Upon cooling, these charges act as a poling field, reinducing polarization.

**Conclusion**

Our work demonstrates the coexistence of superconductivity and ferroelectricity at LAO/KTO interfaces, offering a unique platform for studying their interplay. The switchable polarization modulates the interfacial potential well by reducing ε and altering built-in field, driving $T_c$ enhancement and bistable transport properties. These findings deepen our understanding of the mechanisms coupling superconductivity and ferroelectricity, and open new possibilities for designing multifunctional quantum devices.

# Methods

## Sample fabrication

LAO/KTO interface samples were fabricated by depositing amorphous LAO films onto 0.5 mm thick KTO single-crystalline substrates using pulsed laser deposition (PLD). A 248-nm KrF excimer laser was employed with a laser fluence of 0.7 J/cm$^2$ and a repetition rate of 10 Hz. A single-crystalline LAO target was used. For typical samples, 7-20 nm thick LAO films were grown at 300 °C in an atmosphere of $1\times10^{-5}$ mbar O$_2$ and $1\times10^{-7}$ mbar water vapor[13]. After deposition, the samples were cooled to room temperature under the same atmospheric conditions. In addition, "non-typical" samples were prepared by depositing a 2nm "typical" LAO layer, followed by a 20 nm LAO layer grown at room temperature. The LAO films are highly insulating, with conduction confined to a thin (~5-10 nm thick) KTO layer adjacent to the interface.

## Hall bar devices

Hall bar structures were patterned onto KTO substrates using standard optical lithography and lift-off techniques, with ~200 nm thick AlO$_x$ films serving as a hard mask[32]. The AlO$_x$ films were deposited by PLD at room temperature under base pressure, with a laser fluence of 2.5 J/cm$^2$. To ensure high insulation in the AlO$_x$-covered areas, the patterned substrates were annealed at 300 °C for 2 hours in a flow of 1 bar O$_2$.

Subsequently, LAO films were deposited onto these pre-patterned KTO substrates, forming conducting LAO/KTO interfaces exclusively in the uncovered regions. Each sample contained four identical Hall bar devices. Across different samples, the central Hall bar bridges were fabricated in two sizes: 20 μm in width and 100 μm in length, or 100 μm in width and 500 μm in length.

## Electrical contact

Electrical contacts to the conducting LAO/KTO interfaces were established using ultrasonic bonding with Al wires.

**Gating and poling**

As illustrated in Fig. 1a, a back-gating voltage ($V_G$) was applied between the conducting interface and the bottom silver electrode, with the polarity defined relative to the interface. The value of $V_G$ represents the bias applied to the bottom silver electrode. Throughout the gating process, the leakage current was consistently below 10 nA.

The process of applying $V_G$ and subsequently setting it to 0 is referred to as "poling", similar to operations in conventional ferroelectrics. The states "$0^+$" and "$0^-$" represent the $V_G = 0$ state after removing positive and negative $V_G$, respectively. For clarity, we also denote the $V_G = 0$ state after poling with $V_G$ as $V_G^{0\pm}$. For example, "$120^{0+}$" indicates the $V_G = 0$ state following an applied $V_G = +120$ V. This nomenclature is extended to all cases accordingly.

**Transport measurements**

Low-temperature transport measurements were performed using a commercial $^4$He cryostat equipped with a $^3$He insert (Cryogenic Ltd.). A four-probe DC technique was employed, utilizing a Keithley 6221 current source and a Keithley 2182A nanovoltmeter. Carrier mobility ($\mu$) and density ($n_s$) were obtained from Hall effect measurements conducted on Hall bar devices.

# Acknowledgements

This work was supported by the National Key R&D Program of China (2023YFA1406400), National Natural Science Foundation of China (Grants No. 12325402 and 11934016), Innovation Program for Quantum Science and Technology (Grant No.2021ZD0300200), and the Key R&D Program of Zhejiang Province, China (Grants No. 2020C01019 and 2021C01002).

## Author contributions

Y.X. and M.Z. conceived the study and proposed the strategy. M.Z., M.Q. and Y.S. prepared the samples. S.H. contributed to the development of the electrostatic measurements. M.Z., M.Q. and Y.S. conducted transport measurements and performed analysis. All authors participated in the discussion on the paper. M.Z. and Y.X. wrote the manuscript with input from all the authors.

## References

1. Anderson, P. W. & Blount, E. I. Symmetry considerations on martensitic transformations: 'ferroelectric' metals? *Phys. Rev. Lett.* **14**, 217–219 (1965).

2. Shi, Y. *et al.* A ferroelectric-like structural transition in a metal. *Nat. Mater.* **12**, 1024–1027 (2013).

3. Zhou, W. X. & Ariando, A. Review on ferroelectric/polar metals. *Jpn. J. Appl. Phys.* **59**, SI0802 (2020).

4. Amon, A. *et al.* Noncentrosymmetric superconductor BeAu. *Phys. Rev. B* **97**, 014501 (2018).

5. Salmani-Rezaie, S., Ahadi, K. & Stemmer, S. Polar nanodomains in a ferroelectric superconductor. *Nano Lett.* **20**, 6542–6547 (2020).

6. Jindal, A. *et al.* Coupled ferroelectricity and superconductivity in bilayer $T_d$-MoTe$_2$. *Nature* **613**, 48–52 (2023).

7. Ohtomo, A. & Hwang, H. Y. A high-mobility electron gas at the LaAlO$_3$/SrTiO$_3$ heterointerface. *Nature* **427**, 423–426 (2004).

8. Reyren, N. *et al.* Superconducting interfaces between insulating oxides. *Science* **317**, 1196–1199 (2007).

9. Caviglia, A. D. *et al.* Electric field control of the LaAlO$_3$/SrTiO$_3$ interface ground state. *Nature* **456**, 624–627 (2008).

10. King, P. D. C. *et al.* Subband structure of a two-dimensional electron gas formed at the polar surface of the strong spin-orbit perovskite KTaO$_3$. *Phys. Rev. Lett.* **108**, 117602 (2012).

11. Liu, C. *et al.* Two-dimensional superconductivity and anisotropic transport at KTaO$_3$ (111) interfaces. *Science* **371**, 716–721 (2021).

12. Chen, Z. *et al.* Two-dimensional superconductivity at the LaAlO$_3$/KTaO$_3$ (110) heterointerface. *Phys. Rev. Lett.* **126**, 026802 (2021).

13. Chen, Z. *et al.* Electric field control of superconductivity at the LaAlO$_3$/KTaO$_3$ (111) interface. *Science* **372**, 721–724 (2021).

14. Mallik, S. *et al.* Superfluid stiffness of a KTaO$_3$-based two-dimensional electron gas. *Nat. Commun.* **13**, 4625 (2022).

15. Hua, X. *et al.* Superconducting stripes induced by ferromagnetic proximity in an oxide heterostructure. *Nat. Phys.* **20**, 957–963 (2024).

16. Müller, K. A. & Burkard, H. SrTiO$_3$: An intrinsic quantum paraelectric below 4 K. *Phys. Rev. B* **19**, 3593–3602 (1979).

17. Ichikawa, Y., Nagai, M. & Tanaka, K. Direct observation of the soft-mode dispersion in the incipient ferroelectric KTaO$_3$. *Phys. Rev. B* **71**, 092106 (2005).

18. Schooley, J. F., Hosler, W. R. & Cohen, M. L. Superconductivity in semiconducting SrTiO$_3$. *Phys. Rev. Lett.* **12**, 474–475 (1964).

19. Scheerer, G. *et al.* Ferroelectricity, superconductivity, and SrTiO$_3$—passions of K.A. Müller. *Condens. Matter* **5**, 60 (2020).

20. Bussmann-Holder, A., Simon, A. & Büttner, H. Possibility of a common origin to ferroelectricity and superconductivity in oxides. *Phys. Rev. B* **39**, 207–214 (1989).

21. Edge, J. M., Kedem, Y., Aschauer, U., Spaldin, N. A. & Balatsky, A. V. Quantum critical origin of the superconducting dome in SrTiO$_3$. *Phys. Rev. Lett.* **115**, 247002 (2015).

22. Stucky, A. *et al.* Isotope effect in superconducting n-doped SrTiO$_3$. *Sci. Rep.* **6**, 37582 (2016).

23. Tomioka, Y., Shirakawa, N. & Inoue, I. H. Superconductivity enhancement in polar metal regions of Sr$_{0.95}$Ba$_{0.05}$TiO$_3$ and Sr$_{0.985}$Ca$_{0.015}$TiO$_3$ revealed by systematic Nb doping. *npj Quantum Mater.* **7**, 1–8 (2022).

24. Tomioka, Y., Shirakawa, N., Shibuya, K. & Inoue, I. H. Enhanced superconductivity close to a non-magnetic quantum critical point in electron-doped strontium titanate. *Nat. Commun.* **10**, 738 (2019).

25. Herrera, C. *et al.* Strain-engineered interaction of quantum polar and superconducting phases. *Phys. Rev. Mater.* **3**, 124801 (2019).

26. Ahadi, K. *et al.* Enhancing superconductivity in $SrTiO_3$ films with strain. *Sci. Adv.* **5**, eaaw0120 (2019).

27. Russell, R. *et al.* Ferroelectric enhancement of superconductivity in compressively strained $SrTiO_3$ films. *Phys. Rev. Mater.* **3**, 091401 (2019).

28. Rischau, C. W. *et al.* A ferroelectric quantum phase transition inside the superconducting dome of $Sr_{1-x}Ca_xTiO_{3-\delta}$. *Nat. Phys.* **13**, 643–648 (2017).

29. Bréhin, J. *et al.* Switchable two-dimensional electron gas based on ferroelectric Ca: $SrTiO_3$. *Phys. Rev. Mater.* **4**, 041002 (2020).

30. Thompson, J. R., Boatner, L. A. & Thomson, J. O. Very low-temperature search for superconductivity in semiconducting $KTaO_3$. *J. Low Temp. Phys.* **47**, 467–475 (1982).

31. Ueno, K. *et al.* Discovery of superconductivity in $KTaO_3$ by electrostatic carrier doping. *Nat. Nanotechnol.* **6**, 408–12 (2011).

32. Ren, T. *et al.* Two-dimensional superconductivity at the surfaces of $KTaO_3$ gated with ionic liquid. *Sci. Adv.* **8**, eabn4273 (2022).

33. Hua, X. *et al.* Tunable two-dimensional superconductivity and spin-orbit coupling at the $EuO/KTaO_3(110)$ interface. *npj Quantum Mater.* **7**, 97 (2022).

34. Liu, C. *et al.* Tunable superconductivity and its origin at $KTaO_3$ interfaces. *Nat. Commun.* **14**, 951 (2023).

35. Qiao, W. *et al.* Gate tunability of the superconducting state at the $EuO/KTaO_3$ (111) interface. *Phys. Rev. B* **104**, 184505 (2021).

36. Fujii, Y. & Sakudo, T. Dielectric and optical properties of $KTaO_3$. *J. Phys. Soc. Jpn.* **41**, 888–893 (1976).

37. Aktas, O., Crossley, S., Carpenter, M. A. & Salje, E. K. H. Polar correlations and defect-induced ferroelectricity in cryogenic $KTaO_3$. *Phys. Rev. B* **90**, 165309 (2014).

38. Ang, C. & Yu, Z. dc electric-field dependence of the dielectric constant in polar dielectrics: Multipolarization mechanism model. *Phys. Rev. B* **69**, 174109 (2004).

39. DiAntonio, P., Vugmeister, B. E., Toulouse, J. & Boatner, L. A. Polar fluctuations and first-order Raman scattering in highly polarizable $KTaO_3$ crystals with off-center Li and

Nb ions. *Phys. Rev. B* **47**, 5629–5637 (1993).

40. Mota, D. A. *et al.* Induced polarized state in intentionally grown oxygen deficient $KTaO_3$ thin films. *J. Appl. Phys.* **114**, 034101 (2013).

41. Tyunina, M. *et al.* Evidence for strain-induced ferroelectric order in epitaxial thin-film $KTaO_3$. *Phys. Rev. Lett.* **104**, 227601 (2010).

42. Trybuła, Z., Miga, S., Łoś, S., Trybuła, M. & Dec, J. Evidence of polar nanoregions in quantum paraelectric $KTaO_3$. *Solid State Commun.* **209–210**, 23–26 (2015).

43. Uwe, H., Lyons, K. B., Carter, H. L. & Fleury, P. A. Ferroelectric microregions and Raman scattering in $KTaO_3$. *Phys. Rev. B* **33**, 6436–6440 (1986).

44. Warren, W. L., Vanheusden, K., Dimos, D., Pike, G. E. & Tuttle, B. A. Oxygen vacancy motion in perovskite oxides. *J. Am. Ceram.* **79**, 536–538 (1996).

45. Huang, G. *et al.* Time-dependent resistance of quasi-two-dimensional electron gas on $KTaO_3$. *Appl. Phys. Lett.* **117**, 171603 (2020).

46. Nukala, P. *et al.* Reversible oxygen migration and phase transitions in hafnia-based ferroelectric devices. *Science* **372**, 630–635 (2021).

47. Chen, Z. *et al.* Dual-gate modulation of carrier density and disorder in an oxide two-dimensional electron system. *Nano Lett.* **16**, 6130–6136 (2016).

48. Chen, Z. *et al.* Carrier density and disorder tuned superconductor-metal transition in a two-dimensional electron system. *Nat. Commun.* **9**, 4008 (2018).

49. Minohara, M. *et al.* Dielectric collapse at the $LaAlO_3$/$SrTiO_3$ (001) heterointerface under applied electric field. *Sci. Rep.* **7**, 9516 (2017).

50. Peelaers, H. *et al.* Impact of electric-field dependent dielectric constants on two-dimensional electron gases in complex oxides. *Appl. Phys. Lett.* **107**, 183505 (2015).

51. Copie, O. *et al.* Towards two-dimensional metallic behavior at $LaAlO_3$/$SrTiO_3$ interfaces. *Phys. Rev. Lett.* **102**, 216804 (2009).

52. Hong, S., Nakhmanson, S. M. & Fong, D. D. Screening mechanisms at polar oxide heterointerfaces. *Rep. Prog. Phys.* **79**, 076501 (2016).

53. Kalinin, S. V., Kim, Y., Fong, D. D. & Morozovska, A. N. Surface-screening mechanisms in ferroelectric thin films and their effect on polarization dynamics and domain structures. *Rep. Prog. Phys.* **81**, 036502 (2018).

# Figures

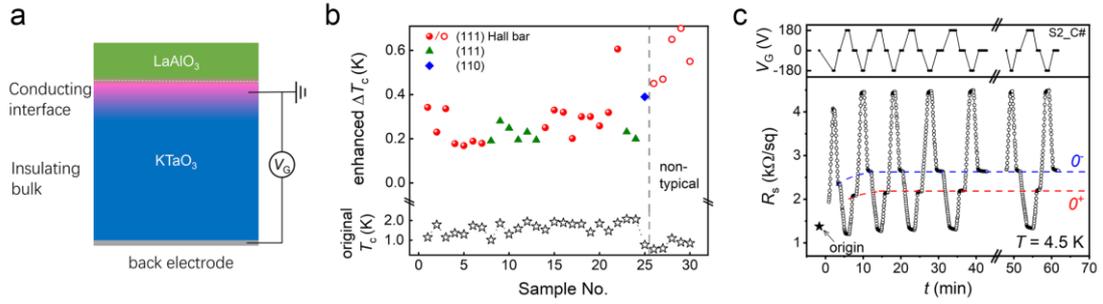

**Fig. 1 | Universally enhanced superconductivity and bistability. a,** Schematic illustration of the gating setup for the LAO/KTO interface, with $V_G$ applied across the KTO substrate. The polarity of $V_G$ is defined relative to the interface. "Poling" refers to the process of applying $V_G$ and then removing it. **b,** Enhancement of $T_c$ ($\Delta T_c$) for multiple LAO/KTO samples after poling with $V_G = -180$ V. Open black stars: original $T_c$. Closed blue diamonds: unpatterned LAO/KTO(110) samples. Closed green triangles: unpatterned LAO/KTO(111) samples. Closed red circles: LAO/KTO(111) Hall bar devices. Open red circles: LAO/KTO(111) Hall bar devices where LAO films were deposited using a non-typical procedure (details provided in **Methods**). **c,** Time-dependent sheet resistance ($R_s$) at $T = 4.5$ K measured while $V_G$ was switched repeatedly between 0, -180, 0, +180 V. "Origin" denotes the state before any $V_G$ was applied. At $V_G = 0$, two distinct $R_s$ states are observed: "$0^+$" and "$0^-$", corresponding to the states after removing positive or negative $V_G$, respectively.

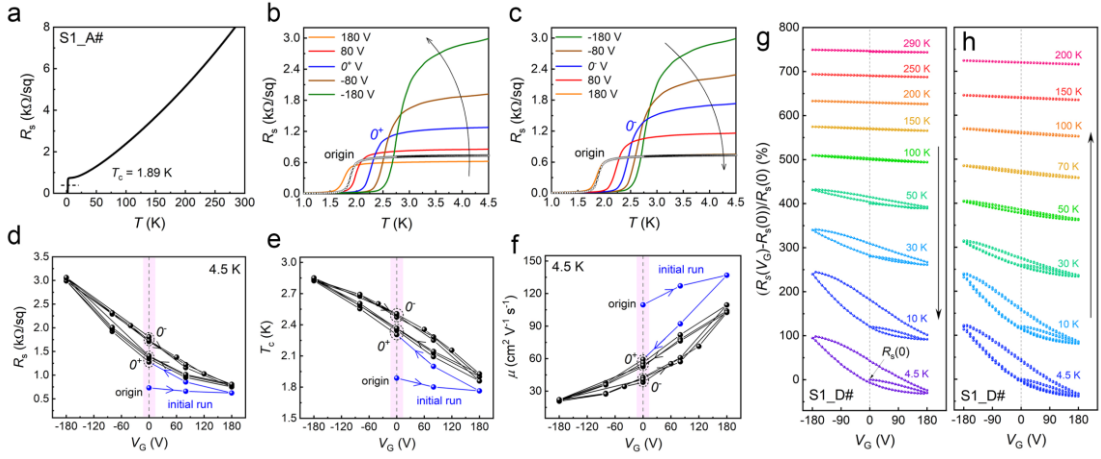

**Fig. 2 | Ferroelectric hysteresis at the LAO/KTO interface.** Consecutive gating cycles between $V_G$ = +180 V and -180 V were performed on a typical LAO/KTO(111) Hall bar device (S1_A#): **a,** Temperature-dependent $R_s(T)$ curve for the "origin" state (before any gating). **b, c,** Temperature-dependent $R_s(T)$ curves during a single gating cycle: **b,** sweeping $V_G$ from +180 V to -180 V; **c,** sweeping $V_G$ from -180 V to +180 V. **d, e, f,** Hysteresis loops during consecutive gating cycles: **d,** $R_s$-$V_G$; **e,** $T_c$-$V_G$; **f,** $\mu$-$V_G$. The pink shading highlights the $V_G$ = 0 region. **g, h,** Evolution of $R_s$-$V_G$ loops with temperature: **g,** measured in decreasing temperature order; **h,** measured in increasing temperature order. To improve clarity, the normalized form [$R_s(V_G)$-$R_s(0)$]/$R_s(0)$ is used, with curves shifted vertically for better visualization.

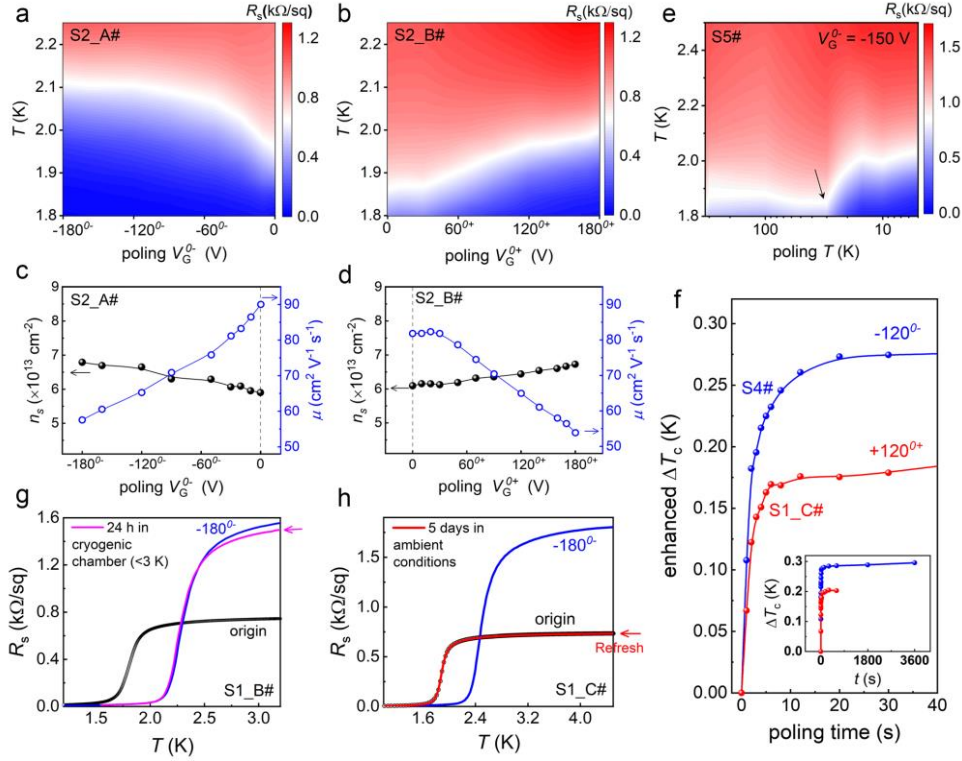

**Fig. 3 | Effects of poling $V_G$, temperature, time, and stability. a, b,** Temperature-dependent $R_s(T)$ maps as a function of negative ($V_G < 0$) and positive ($V_G > 0$) poling $V_G$, respectively. **c, d,** Corresponding carrier density ($n_s$) and mobility ($\mu$) measured at $T = 4.5$ K. **e,** Temperature-dependent $R_s(T)$ map as a function of poling temperature, using a fixed poling $V_G = -150$ V. **f,** Evolution of the enhanced $\Delta T_c$ with cumulative poling time, performed at $T = 4.5$ K with poling $V_G = \pm 120$ V on two different samples. The inset shows the same data on an extended time scale. **g,** The poling-induced state remains non-volatile at low temperatures. **h,** The poling-induced state recovers to the "origin" state after being left in ambient conditions for several days. For each experiment, fresh LAO/KTO samples were used as needed to ensure measurements began from the "origin" state.

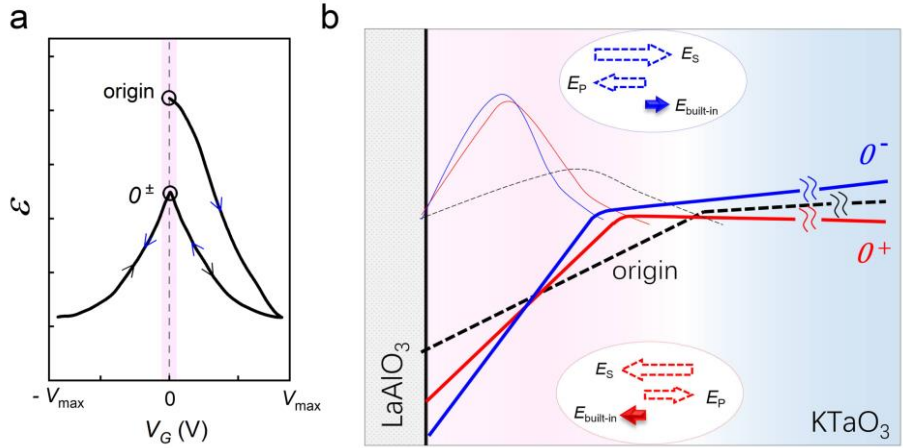

**Fig. 4 | Effect of ferroelectric polarization on the interfacial potential well. a,** Schematic evolution of the dielectric constant ($\varepsilon$) of KTO under an applied external electric field, reproduced based on experimental results from Ref. [36,38]. The pink shading highlights the "$0^{\pm}$" states, which exhibit a reduced $\varepsilon$ compared to the "origin" state. **b,** Schematic diagrams of the interfacial potential well for the "origin" state (black dashed lines), "$0^-$" state (blue lines), and "$0^+$" state (red lines). The thick lines represent the confinement potential well, while the thin lines represent the electron envelope wavefunction confined within the well. The upper and lower insets illustrate the electric fields contributing to the potential well, arising from the ferroelectric polarization in the interfacial conducting layer. $E_P$, the field from polarization itself; $E_S$, the field from polarization-associated screening charges, $E_{\text{built-in}}$, the sum of $E_P$ and $E_S$, which modulates the potential well profile. Notably, $E_P$ is in the opposite direction to the field induced by $V_G$. After $V_G$ is removed, a residual tuning effect persists due to ferroelectric polarization. This residual effect aligns with the direction of $V_G$, suggesting that $E_S$ opposites and exceeds $E_P$.

**Extended Data Figures**

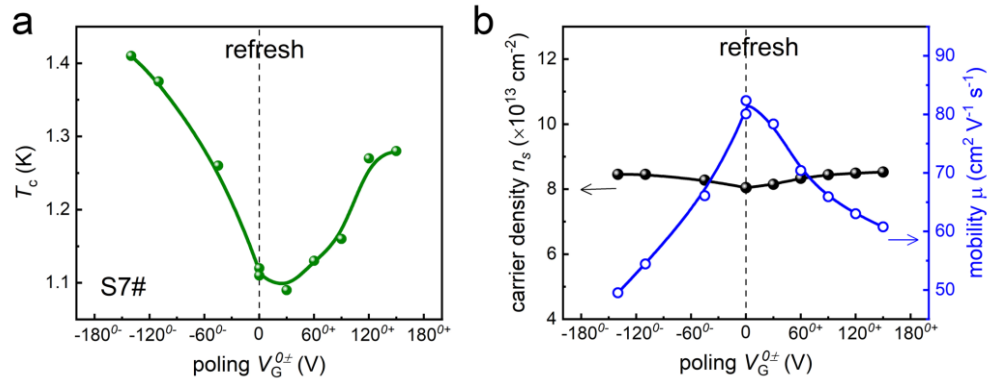

**Extended Data Fig. 1 | Poling a LAO/KTO(111) sample (S7#) with increasing |$V_G$| from 0 to 180 V for both polarities.** The "origin" state was restored through a "refresh" process by leaving the sample in ambient conditions for several days. **a,** Evolution of $T_c$ as a function of poling $V_G$. **b,** Evolutions of carrier density ($n_s$, black closed circles) and mobility ($\mu$, blue open circles) as functions of poling $V_G$.

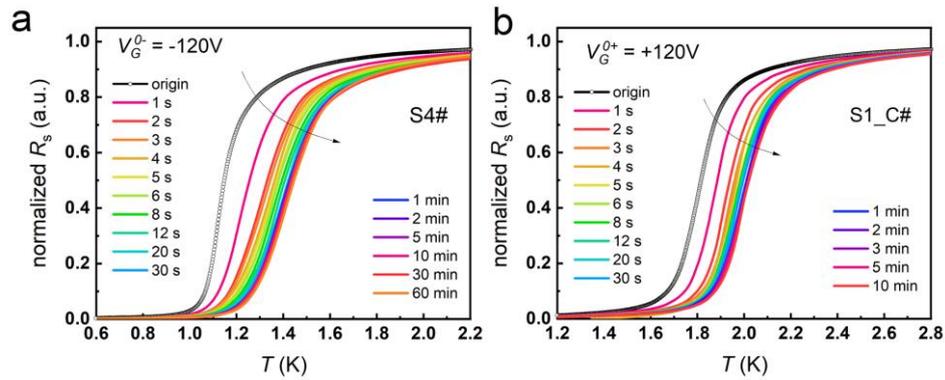

**Extended Data Fig. 2 | Evolution of normalized $R_s(T)$ curves with increasing cumulative poling time.** The poling processes were conducted at $T = 4.5$ K, with cumulative poling times as labeled. **a,** Results for a poling $V_G$ of -120 V. **b,** Results for a poling $V_G$ of +120 V.

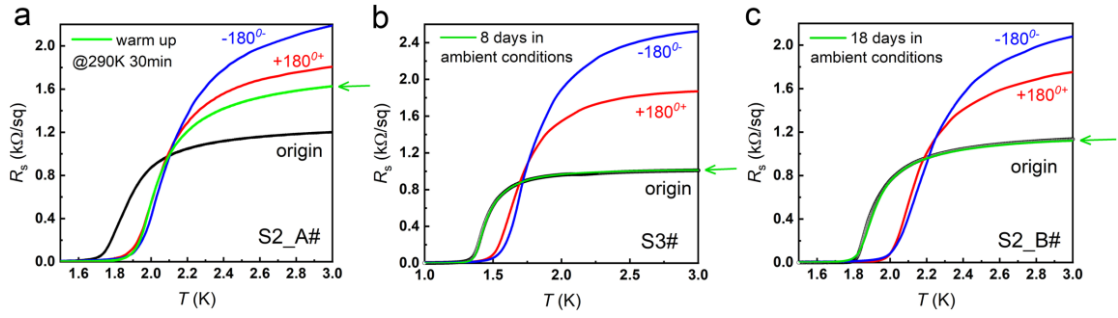

**Extended Data Fig. 3 | Recovery behavior of the poling-induced state. a,** Temperature-dependent $R_s(T)$ curve after warming the sample to 290 K and holding it for 30 minutes in the cryostat chamber; **b,** $R_s(T)$ curve after leaving the sample in ambient conditions for 8 days; **c,** $R_s(T)$ curve after leaving the sample in ambient conditions for 18 days. For each sample, the temperature-dependent $R_s(T)$ curves corresponding to "origin", "+180$^{0+}$" and "-180$^{0-}$" states are shown for comparison.

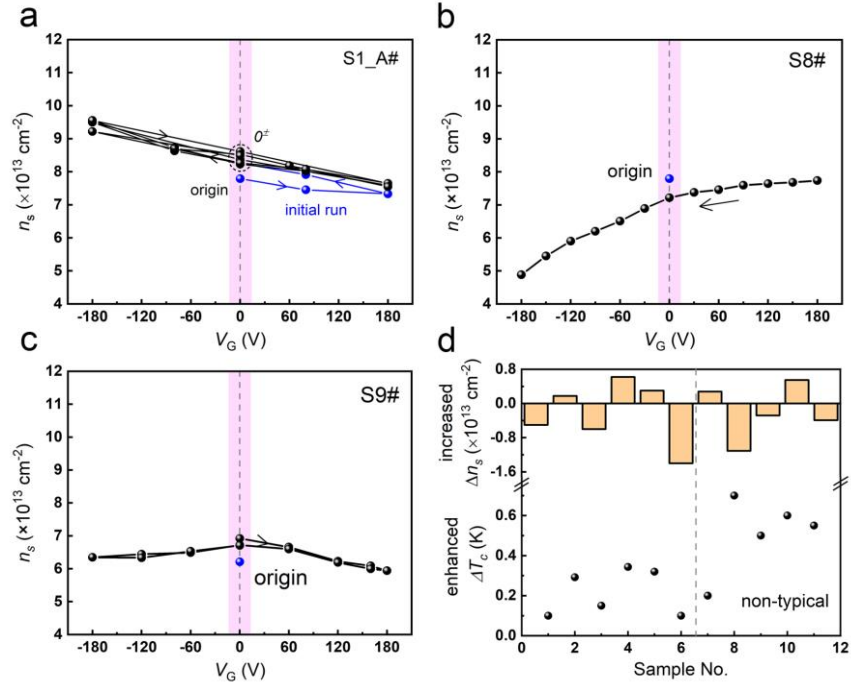

**Extended Data Fig. 4 | Evolution of carrier density ($n_s$) with gating $V_G$ in various LAO/KTO samples.** The pink shading highlights the "$V_G = 0$" region. **a,** Sample S1_A# during the continuous gating cycles described in the main text and Fig. 2. **b,** Sample S8#. **c,** Sample S9#. **d,** Summary of $\Delta n_s$, the difference in $n_s$ between the "$0^{\pm}$" states and their corresponding "origin" state, across various samples. Two key points are noteworthy: (1) As illustrated in panels **a-c**, the $n_s$-$V_G$ relationship varies in different samples, supporting the conclusion that the primary effect of $V_G$ is not on $n_s$ through a simple capacitance mechanism. Instead, $n_s$ is influenced by several factors, the detailed discussion of which is beyond the scope of this study. (2) As shown in panel **d**, while the enhancement in $T_c$ (indicating the presence of ferroelectric polarization) is a universal feature, $\Delta n_s$ can be either positive or negative. This strongly supports the conclusion that the observed phenomena are not driven by changes in $n_s$. The term "non-typical" is used as defined in **Fig. 1b**.